\documentclass[aps,prc,twocolumn,groupedaddress,showpacs]{revtex4}
\usepackage{graphicx}

\begin{document}


\title{The astrophysical rate of $^{15}$O($\alpha,\gamma)^{19}$Ne via the ($p,t$) reaction in inverse kinematics}


\author{B. Davids}
\email[]{davids@kvi.nl}
\author{A. M. van den Berg}
\author{P. Dendooven}
\affiliation{Kernfysisch Versneller Instituut, Zernikelaan 25, 9747 AA Groningen, The Netherlands}
\author{F. Fleurot}
\altaffiliation{Present address: Department of Physics, Laurentian University, Sudbury ON, P3E 2C6, Canada}
\affiliation{Kernfysisch Versneller Instituut, Zernikelaan 25, 9747 AA Groningen, The Netherlands}
\author{M. Hernanz}
\affiliation{Institut d'Estudis Espacials de Catalunya, Ed. Nexus-201, C/Gran Capit\`a 2-4, 08034 Barcelona, Spain}
\author{M. Hunyadi}
\author{M. A. de Huu}
\affiliation{Kernfysisch Versneller Instituut, Zernikelaan 25, 9747 AA Groningen, The Netherlands}
\author{J. Jos\'e}
\affiliation{Institut d'Estudis Espacials de Catalunya, Ed. Nexus-201, C/Gran Capit\`a 2-4, 08034 Barcelona, Spain}
\affiliation{Departament de F\'{\i}sica i Enginyeria Nuclear, Universitat Polit\`ecnica de Catalunya, Av. Victor Balaguer s/n E-08800 Vilanova i la Geltr\'u, Barcelona, Spain}
\author{K. E. Rehm}
\affiliation{Physics Division, Argonne National Laboratory, Argonne IL 60439}
\author{R. E. Segel}
\affiliation{Department of Physics, Northwestern University, Evanston IL 60208}
\author{R. H. Siemssen}
\author{H. W. Wilschut}
\author{H. J. W\"{o}rtche}
\affiliation{Kernfysisch Versneller Instituut, Zernikelaan 25, 9747 AA Groningen, The Netherlands}
\author{A. H. Wuosmaa}
\altaffiliation{Present address: Department of Physics, Western Michigan University, Kalamazoo MI, 49008}
\affiliation{Physics Division, Argonne National Laboratory, Argonne IL 60439}


\date{\today}

\begin{abstract}
A recoil coincidence technique has been applied to measure the $\alpha$-decay branching ratios of near-threshold states in $^{19}$Ne. Populating these states using the ($p,t$) reaction in inverse kinematics, we detected the recoils and their decay products with 100\% geometric efficiency using a magnetic spectrometer. Combining our branching ratio measurements with independent determinations of the radiative widths of these states, we calculate the astrophysical rate of $^{15}$O($\alpha,\gamma)^{19}$Ne. Using this reaction rate, we perform hydrodynamic calculations of nova outbursts and conclude that no significant breakout from the hot CNO cycles into the $rp$ process occurs in novae via $^{15}$O($\alpha,\gamma)^{19}$Ne.
\end{abstract}


\maketitle

\section{Introduction}

Explosive thermonuclear fusion reactions on the surfaces of accreting compact objects in binary star systems cause the astronomical phenomena of novae and x-ray bursts. In novae, which take place on white dwarfs, hydrogen is burned at different stages of the outburst by combinations of the $pp$ chains, the hot and cold CNO cycles, and the NeNa and MgAl cycles \cite{jose98,starrfield00}. The main nuclear activity involves ($p,\gamma$) and ($p,\alpha$) reactions interspersed with $\beta^+$ decays. In the case of massive ONe novae, marginal activity is also seen in nuclei beyond Si, with a likely nucleosynthetic endpoint around Ca \cite{jose01}.  X-ray bursts occur on the surfaces of neutron stars, where high peak temperatures allow significant nucleosynthesis beyond Ca. During these violent thermonuclear runaways, helium is burned via the 3$\alpha$ reaction, and intermediate mass nuclei (F through Sc) are produced through the $\alpha p$ process, a succession of ($\alpha,p$) and ($p,\gamma$) reactions \cite{wallace81}. Heavier nuclei are synthesized via the $rp$ process \cite{wallace81,schatz98}, a sequence of rapid proton captures and $\beta^+$ decays which further processes seed nuclei left by the $\alpha p$ process. Nucleosynthesis in x-ray bursts is limited by a closed SnSbTe cycle \cite{schatz01}. Three reactions potentially important for breakout from the hot CNO cycles have been identified, $^{15}$O($\alpha,\gamma)^{19}$Ne, $^{18}$F($p,\gamma)^{19}$Ne, and $^{18}$Ne($\alpha,p)^{21}$Na \cite{wiescher99}. This paper is concerned with the first of these. Direct measurements of the $^{15}$O($\alpha$,$\gamma$)$^{19}$Ne reaction at astrophysically relevant energies have not yet been performed, as the small cross sections would require very high intensity radioactive $^{15}$O beams. Currently, the best information about the astrophysical reaction rate comes from studies of the decay properties of resonances in $^{19}$Ne that contribute to the reaction at the relevant temperatures of 0.1 - 2 GK.

Calculation of the reaction rate requires the $\alpha$ widths $\Gamma_{\alpha}$ and radiative widths $\Gamma_{\gamma}$ of states in $^{19}$Ne lying just above the $\alpha$-emission threshold at 3.529 MeV. Since these states lie below the proton and neutron separation energies of 6.4 and 11.6 MeV respectively \cite{audi95}, they can only decay by the emission of $\alpha$ particles or $\gamma$ rays. Hence knowledge of either the radiative width or the total width $\Gamma$ of a state plus its $\alpha$-decay branching ratio B$_\alpha$ is sufficient to determine its contribution to the astrophysical reaction rate. In the past, $\alpha$-decay branching ratios for these states have been measured by populating the states in transfer reactions and using solid-state detectors to count either $\alpha$ particles \cite{magnus90,kubono02}, or both $\alpha$ particles and $^{15}$O nuclei \cite{laird02} from the subsequent decays. While these studies provided valuable data on higher-lying states, none yielded useful information on the 3/2$^+$ level at 4.033 MeV which dominates the $^{15}$O($\alpha,\gamma)^{19}$Ne reaction rate at the temperatures found e.g. in novae. Previously, the only experimental information on the contribution made by this state came from measurements of transfer reactions populating the analog state in the mirror nucleus $^{19}$F \cite{mao95,mao96}.

\section{Experimental Method}

\begin{figure}\includegraphics[width=\linewidth]{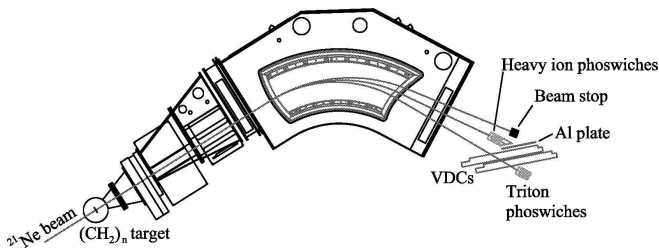} \caption{Experimental setup for the measurement of $\alpha$-decay branching ratios of states in $^{19}$Ne using a recoil coincidence technique at the Big-Bite Spectrometer of the KVI.} \label{fig1} \end{figure}

We have measured the $\alpha$-decay branching ratios of all the states in $^{19}$Ne relevant to the astrophysical rate of the $^{15}$O($\alpha,\gamma)^{19}$Ne reaction \cite{davids03}. This measurement was carried out at the Kernfysisch Versneller Instituut (KVI) using a recoil coincidence technique \cite{rehm00}, through which we have detected both $\alpha$- and $\gamma$-decaying recoils with 100\% geometric efficiency and unambiguous discrimination in a magnetic spectrometer. We populated states in $^{19}$Ne via the $p(^{21}$Ne,$t)^{19}$Ne reaction at an incident beam energy of 43 MeV/u. A 910 MeV $^{21}$Ne beam provided by the variable energy, superconducting cyclotron AGOR bombarded a 1 mg cm$^{-2}$ (CH$_2)_n$ target. Both triton ejectiles and $^{19}$Ne recoils entered the Big-Bite Spectrometer (BBS) \cite{berg95}, which was positioned at 0$^{\circ}$. The $^{19}$Ne recoils subsequently deexcited by the emission of $\gamma$ rays, retaining their identities as $^{19}$Ne, or by the emission of $\alpha$ particles, resulting in the formation of $^{15}$O decay products. Fig.\ \ref{fig1} schematically illustrates the experimental arrangement.

The high geometric efficiency of the experimental setup is a consequence of kinematic forward focusing in this reaction of an energetic, heavy projectile on a light target nucleus. There are two solutions to the kinematics of the reaction at laboratory angles around 0$^{\circ}$, one in which the tritons are emitted forward in the center-of-mass system, and one in which they are emitted backward. We have detected those emitted backward in the center-of-mass system, which have laboratory energies around 14 MeV/u. Tritons emitted at  center-of-mass angles around 180$^{\circ}$ in inverse kinematics correspond to forward-going tritons emitted near 0$^{\circ}$ in the ($p,t$) reaction in normal kinematics. For triton ejectiles emitted at laboratory scattering angles of 4$^{\circ}$ or less in this measurement, the $^{19}$Ne recoils emerge at scattering angles of less than 0.4$^{\circ}$. The impulse delivered to the $^{15}$O decay product in an $\alpha$ decay results in an angular spread about the original $^{19}$Ne trajectory. However, the high incident beam energy and low decay energies of the states studied limit the laboratory scattering angles of the $^{15}$O decay products to 1.5$^{\circ}$, well within the angular acceptance of the spectrometer. Simultaneous detection of both ejectile and recoil is possible in this configuration because the BBS has a momentum acceptance of $\pm$ 9.5\% and a solid angle of 9.2 msr ($\pm$ $1.9^{\circ}$ in the dispersive direction and $\pm$ $4.0^{\circ}$ in the non-dispersive direction).

The $^{19}$Ne recoils and $^{15}$O decay products were identified and stopped in two phoswich detectors \cite{leegte92} placed in a vacuum chamber. These phoswiches provided energy loss, total energy, and timing information. Each detector is made up of a 1 mm layer of fast plastic scintillator (NE102A) coupled to a 5 cm layer of slow plastic scintillator (NE115). Both layers are viewed by a single photomultiplier tube, whose signals are integrated for 40 ns and 400 ns in a charge-to-digital converter. The short integration window yields energy-loss signals from the thin, fast plastic layer while the long window is used to obtain the total energy signals from the thick, stopping, slow plastic layer. The spatial extent of these two detectors (6.5 cm $\times$ 6.5 cm each) was sufficient to provide 100\% acceptance for $^{19}$Ne recoils and $^{15}$O decay products from the excitation of $^{19}$Ne states lying at energies up to 5.5 MeV. This fact was confirmed by direct measurements in the non-dispersive direction of the spectrometer, and by raytracing calculations in the dispersive direction.

Triton ejectiles were identified and stopped in a similar array of six phoswich detectors after passing through two position-sensitive vertical drift chambers (VDCs) \cite{woertche01}, all of which were placed in air. Using the position information from the VDCs, we reconstructed the trajectories of the tritons, allowing determination of their kinetic energies and laboratory scattering angles. Excitation energy resolution of 90 keV full-width-at-half-maximum was obtained via raytracing techniques.

The cross section for fragmentation of the 43 MeV/u $^{21}$Ne beam on the C component of the (CH$_2)_n$ target is far larger than that for the $(p,t)$ reaction, so the experiment required distinguishing a small signal from a large background of fragmentation products. An Al plate positioned just behind the heavy-ion phoswich detectors obscured the high-magnetic-rigidity half of the focal plane and stopped many of the heavy fragmentation products. The beam was stopped in a Faraday cup located just in front of this plate and next to the heavy-ion phoswich array.

\begin{figure}\includegraphics[width=\linewidth]{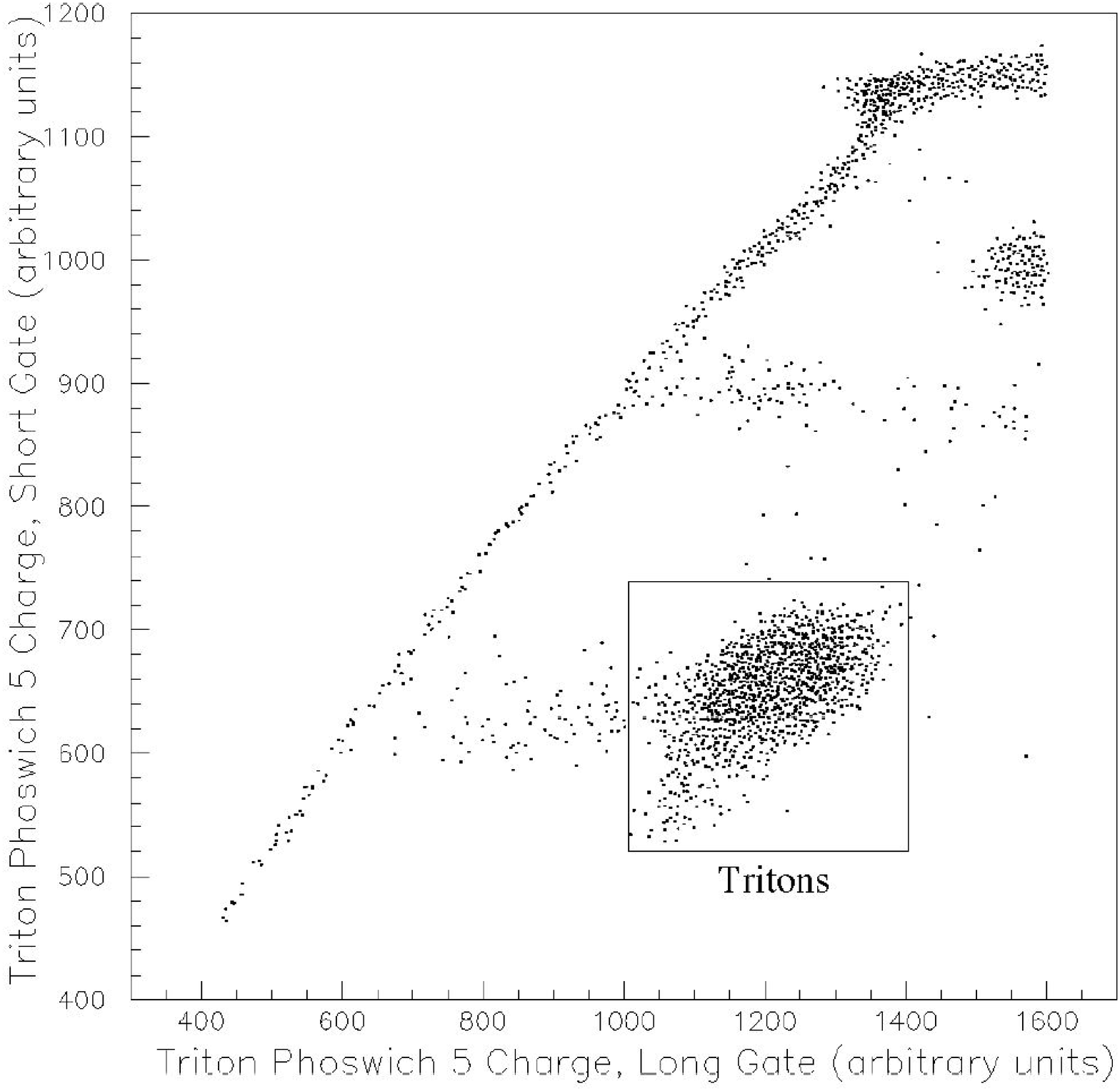} \caption{Triton phoswich integrated charge spectrum. The figure plots the charge collected in a 40 ns window versus that collected during a 400 ns window. The triton group is indicated. Only the low-integrated-charge region is shown here.} \label{fig2} \end{figure}

\begin{figure}\includegraphics[width=\linewidth]{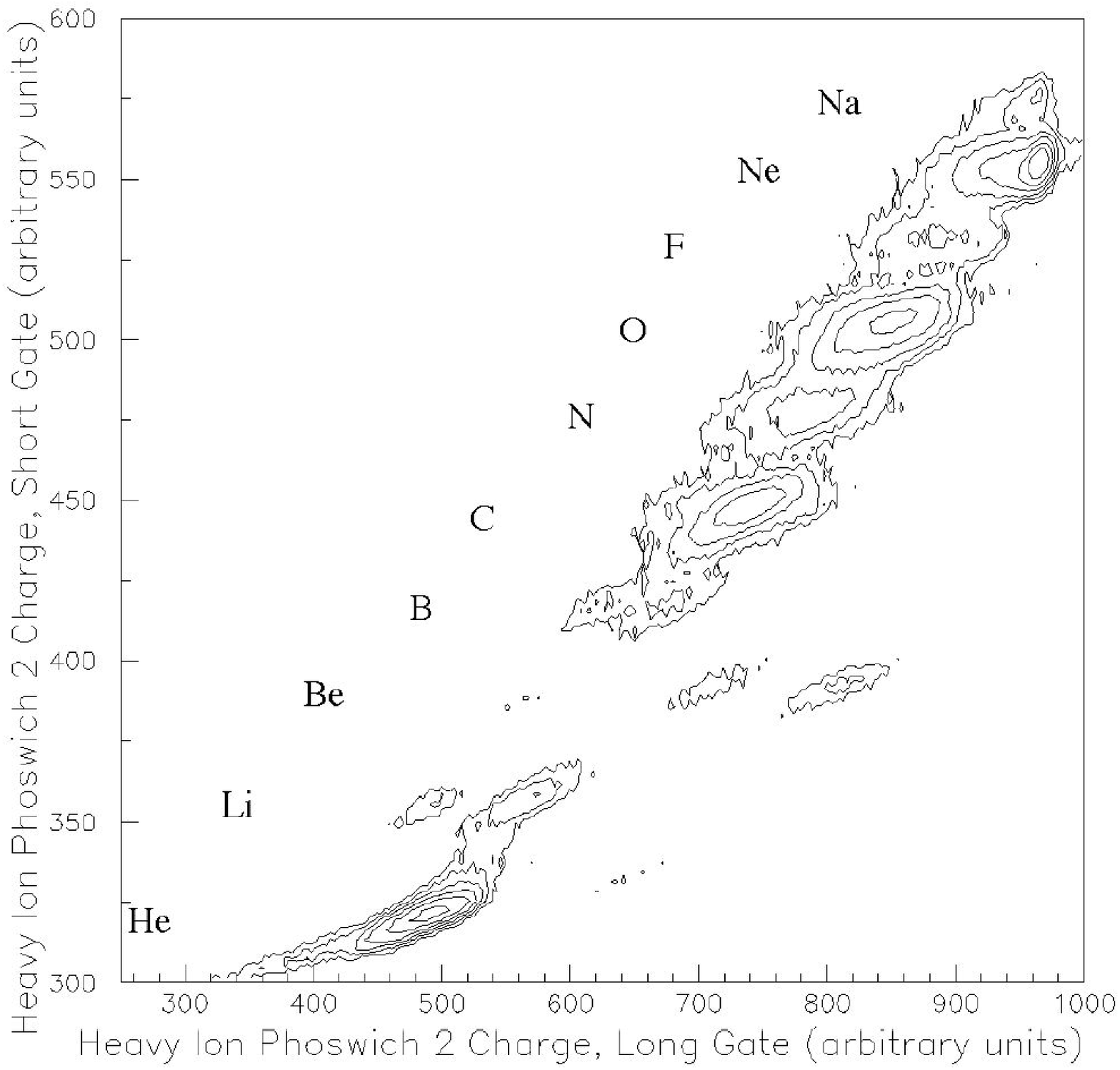} \caption{Heavy-ion phoswich integrated charge spectrum. This contour plot shows the charge collected in a 40 ns window versus that collected during a 400 ns window. The $^{19}$Ne group is centered around (960,550), and the $^{15}$O group about (880,500).} \label{fig3} \end{figure}

\begin{figure}\includegraphics[width=\linewidth]{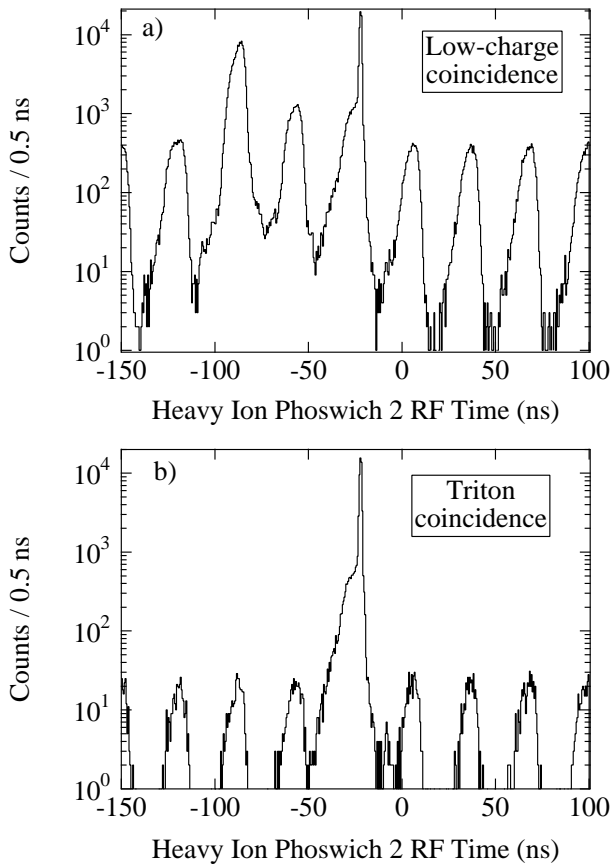} \caption{Heavy-ion phoswich time-of-flight spectrum. Shown is the time of arrival of a particle in a heavy-ion phoswich, in coincidence with a hit in a triton phoswich, measured with respect to the cyclotron radio-frequency signal. Panel (a) shows a TOF spectrum obtained from coincidences with low-integrated-charge hits in the triton phoswiches, while panel (b) shows the same spectrum with a gate on the triton groups in the triton phoswich spectra. The sharp peak near the center is due to $^{19}$Ne coincidences, while the smaller, broader peak underlying it represents $^{15}$O coincidences.} \label{fig4} \end{figure}

Accurate particle identification was essential to this measurement. Fig.\ \ref{fig2} shows the energy loss versus total energy spectrum obtained from one of the triton phoswich detectors. Only the region directly surrounding the triton group is shown. It is clearly separated from the other particle groups. Fig.\ \ref{fig3} shows a similar spectrum obtained from the second of two heavy-ion phoswiches, which was used for detecting both $^{19}$Ne and $^{15}$O nuclei. The separation of the different elements is good throughout, but the isotopic separation decreases with increasing mass and charge. Additional separation is obtained through the use of timing. Fig.\ \ref{fig4} shows the time-of-flight (TOF) spectrum of the second heavy-ion phoswich with respect to the cyclotron radio frequency (RF) signal. Panel (a) shows the TOF of ions arriving in coincidence with low-integrated-charge hits in the triton phoswiches, while panel (b) shows the TOF of ions arriving in coincidence with tritons in the triton phoswiches. The narrow peak evident there is due to triton-$^{19}$Ne coincidences, while the broad peak underlying it is due to triton-$^{15}$O coincidences. Combining such TOF, energy loss, and total energy information we have unambiguously identified coincidences of tritons with both $^{19}$Ne and $^{15}$O nuclei.

\begin{figure}\includegraphics[width=\linewidth]{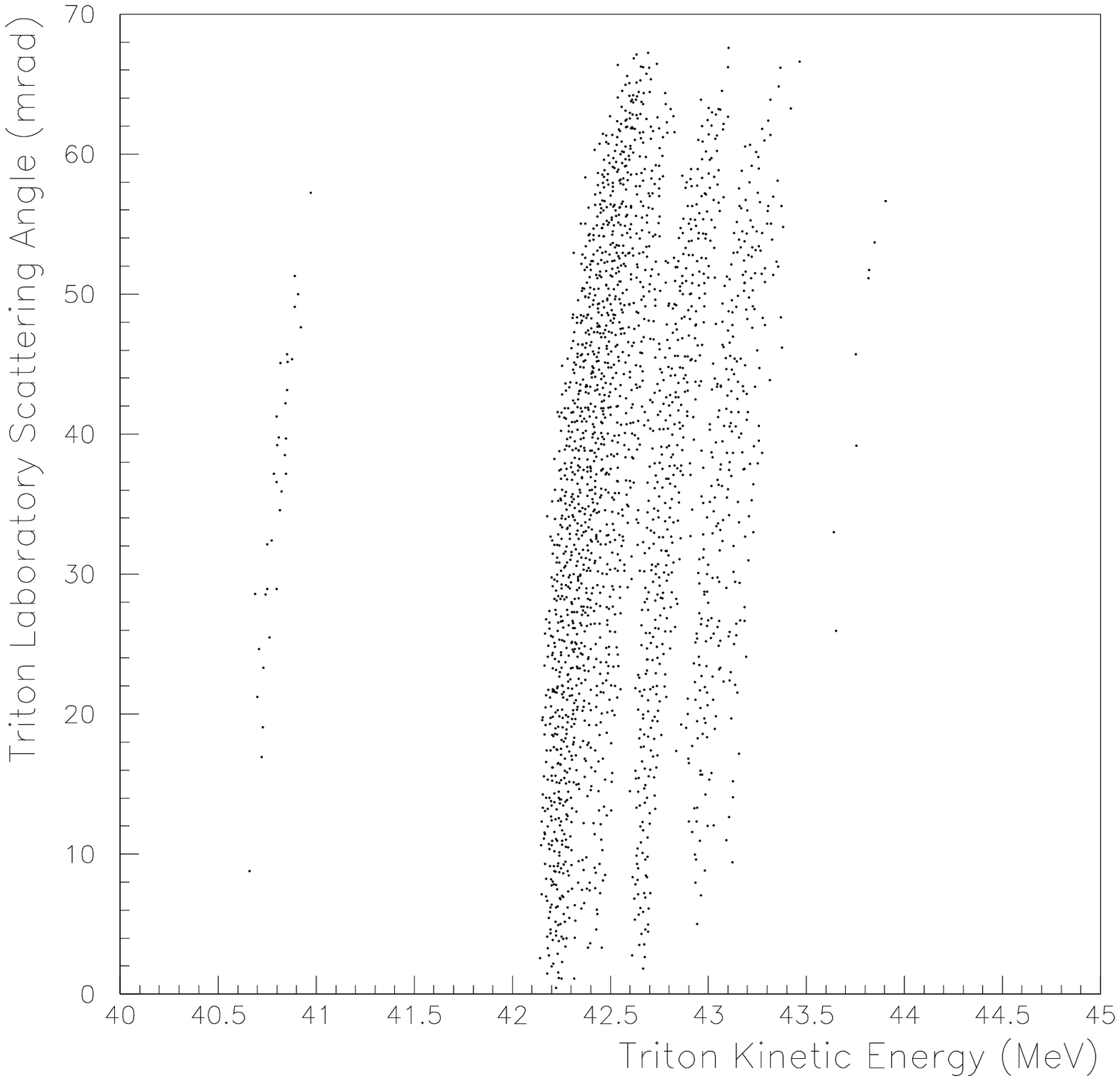} \caption{Triton laboratory scattering angle vs. kinetic energy from $^{19}$Ne-triton coincidences, in logarithmic scale. The curved loci represent $\gamma$ decays of states in $^{19}$Ne.} \label{fig5} \end{figure}

\begin{figure}\includegraphics[width=\linewidth]{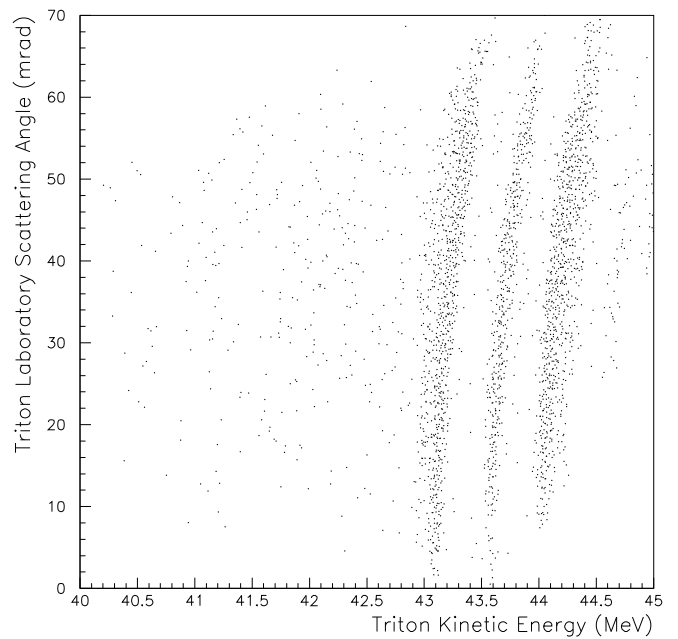} \caption{Triton laboratory scattering angle vs. kinetic energy from $^{15}$O-triton coincidences, in logarithmic scale. The curved loci represent $\alpha$ decays of states in $^{19}$Ne.} \label{fig6} \end{figure}

\section{Results}

Measurements of the scattering angle and kinetic energy of the triton ejectile are sufficient to determine the excitation energy of the $^{19}$Ne recoil in this two-body reaction. Fig.\ \ref{fig5} shows a scatter plot of these quantities from $^{19}$Ne-triton coincidence events, while Fig.\ \ref{fig6} shows the same observables from $^{15}$O-triton coincidences. The curved loci in these figures represent states in $^{19}$Ne, the excitation energies of which increase with increasing triton kinetic energy. This energy dependence and the fact that for a given state the observed triton kinetic energy increases with increasing scattering angle are opposite to the trends usually observed simply because the tritons were emitted backward in the center-of-mass system.

\begin{figure}\includegraphics[width=\linewidth]{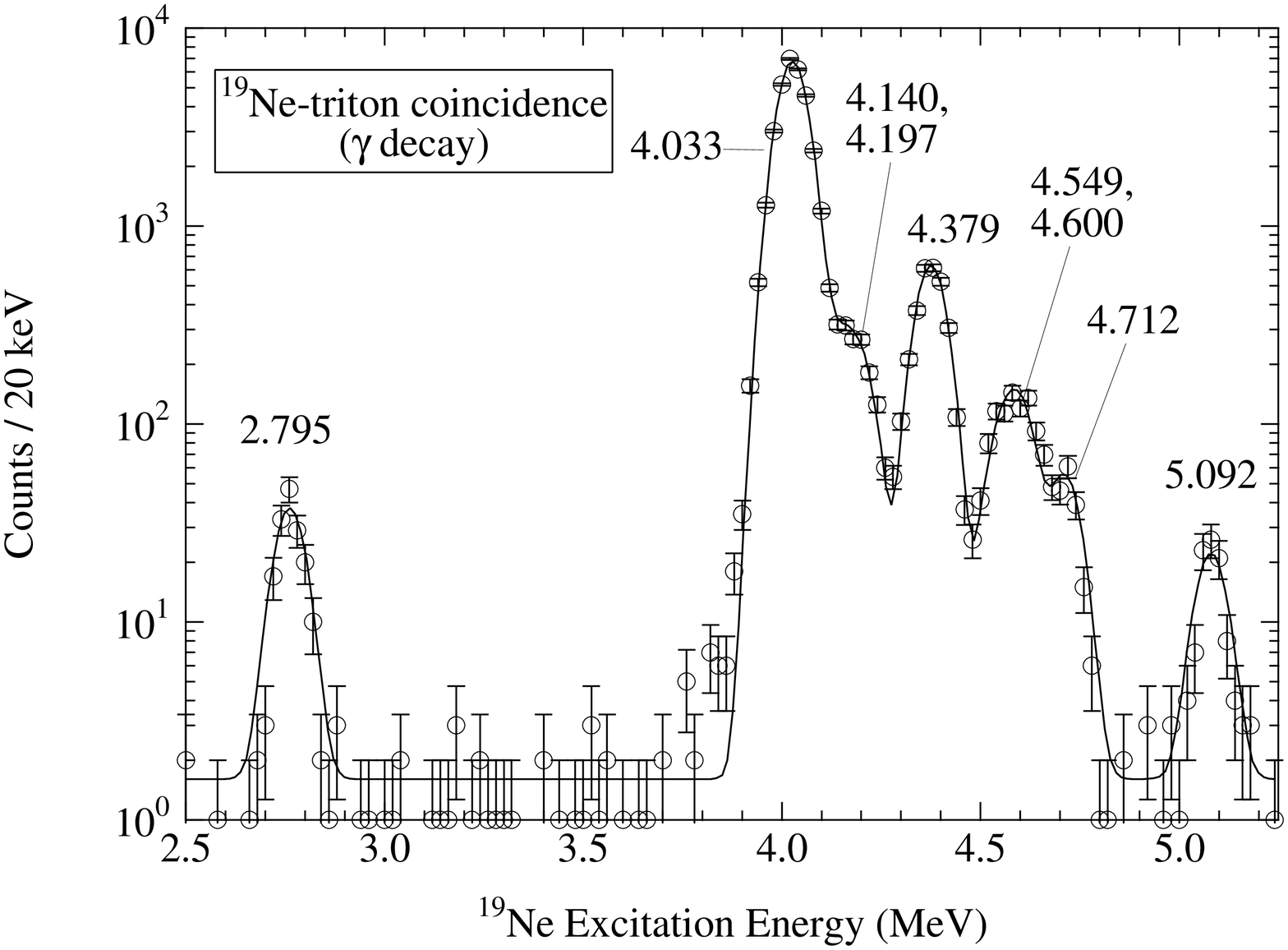} \caption{$^{19}$Ne excitation energy spectrum obtained from $^{19}$Ne-triton coincidences, representing $\gamma$ decays of states in $^{19}$Ne. No $\gamma$ decays of states above 5.092 MeV were observed.} \label{fig7} \end{figure}

\begin{figure}\includegraphics[width=\linewidth]{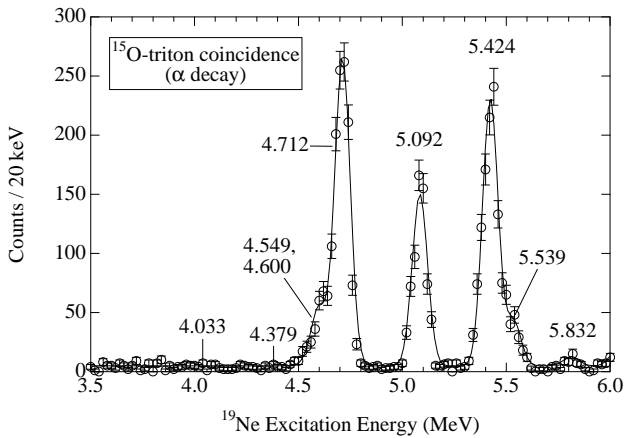} \caption{$^{19}$Ne excitation energy spectrum obtained from $^{15}$O-triton coincidences, representing $\alpha$ decays of states in $^{19}$Ne. There is no statistically significant evidence for $\alpha$ decays from states lying below 4.549 MeV. The $\alpha$-decay threshold lies at 3.529 MeV.} \label{fig8} \end{figure}

\begin{figure}\includegraphics[width=\linewidth]{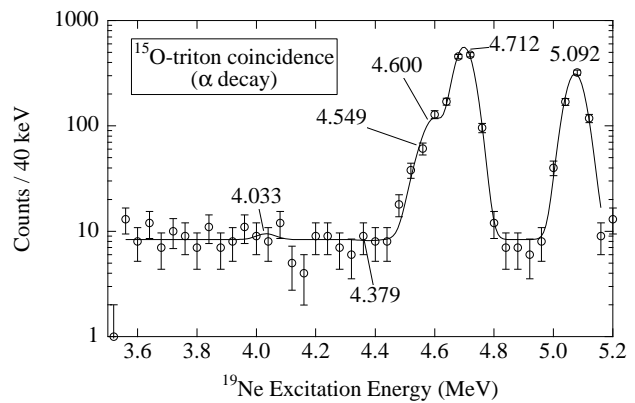} \caption{$^{19}$Ne excitation energy spectrum obtained from $^{15}$O-triton coincidences in logarithmic scale, focusing on the region of astrophysical interest. The curve is the sum of a constant background and six Gaussians corresponding to known states in $^{19}$Ne, the widths of which were fixed by the experimental resolution.} \label{fig9} \end{figure}

Fig.\ \ref{fig7} shows the $^{19}$Ne excitation energy spectrum obtained from $^{19}$Ne-triton coincidences, while Fig.\ \ref{fig8} and Fig.\ \ref{fig9} show that obtained from $^{15}$O-triton coincidences. The solid curves are the sums of constant backgrounds and Gaussians centered at the energies of known states in $^{19}$Ne \cite{tilley95}. The widths of the Gaussians are fixed by the experimental resolution, which is insufficient to completely resolve some pairs of closely-spaced states from one another. In these cases, the yield to each state can be determined from the fit. Both the 4.033 MeV and 4.379 MeV states, which are potentially the most important for the astrophysical reaction rate in novae, are well resolved. The 3/2$^+$, 4.033 MeV state, whose dominant shell-model configuration is (sd)$^5$ (1p)$^{-2}$ \cite{fortune78}, was copiously populated via an s-wave transfer of two neutrons from the 3/2$^+$ ground state of the $^{21}$Ne beam to the proton target, as is evident from Fig.\ \ref{fig7}. The background represents a larger fraction of the total events in the $^{15}$O-triton coincidence spectrum than in the $^{19}$Ne-triton coincidence spectrum, though it is still quite small. A single fragmentation reaction of $^{21}$Ne on a C nucleus in the target can produce $^{15}$O and a triton, whereas it can not produce $^{19}$Ne and a triton. This accounts for the larger background in the $\alpha$-decay spectrum compared with the $\gamma$-decay spectrum, where the background comes from random coincidences between different fragmentation events. The fact that the BBS was set to detect $\sim14$ MeV/u tritons meant that only tritons from the extreme low-energy tail of the triton fragmentation distribution were present in the background.

Of the nine $\gamma$-decaying states indicated in Fig.\ \ref{fig7}, only the four highest-lying levels can also be seen in the $\alpha$-decay spectrum, Fig.\ \ref{fig9}. The reason for this is that the decay energy of the lower-lying states is insufficient to permit appreciable transmission of $\alpha$ particles through the Coulomb and centrifugal barriers. No $\alpha$ decay from the 4.140 MeV and 4.197 MeV states was anticipated, as these decays are hindered by $\ell=4$ centrifugal barriers and low decay energies. As a result, no information on their $\alpha$-decay branching ratios is given here. For the four highest-lying states, at excitation energies of 4.549-, 4.600-, 4.712-, and 5.092 MeV, B$_\alpha$ was determined from the fits.

A different procedure was required for the two astrophysically important states lying at 4.033- and 4.379 MeV, from which no evidence of $\alpha$ decay was immediately apparent. The $\alpha$- and $\gamma$-decay spectra and their associated backgrounds were integrated in 100 keV bins centered at the known energies of the states. A Bayesian statistical analysis \cite{hagiwara02} was applied to these data to obtain upper limits on B$_\alpha$ at both the 90\% and the 99.73\% confidence levels.

Table \ref{table1} contains our measured branching ratios, along with the results of Refs.\ \cite{magnus90,laird02}. Central values and 1$\sigma$ uncertainties are given where they could be obtained; upper limits are specified at the 90\% confidence level. Uncertainties in the present branching ratio determinations are purely statistical. Where more than one measurement is available, a weighted average of B$_\alpha$ is also shown. An exception is made for the 4.379 MeV state because the B$_\alpha$ determination of Ref.\ \cite{magnus90} was based on low statistics, and would more appropriately have been given as an upper limit. Since the present upper limit is more restrictive, it is adopted here. As no evidence for $\gamma$ decay of states lying above 5.092 MeV was observed, the observation of $\alpha$ decay from states at 5.424-, 5.539-, and 5.832 MeV is consistent with 100\% $\alpha$-decay branching ratios for these states.

\begin{table*}
 \caption{\label{table1}Branching ratios B$_{\alpha}\equiv\Gamma_{\alpha}/\Gamma$ and decay widths. Upper limits are specified at the 90\% confidence level.}
 \begin{ruledtabular}
 \begin{tabular}{ccccccccc}
E$_{x}$ (MeV)&J$^{\pi}$&B$_{\alpha}$ (present work)&B$_{\alpha}$ (Ref.\ \cite{magnus90}) & B$_{\alpha}$ (Ref.\ \cite{laird02})&B$_{\alpha}$ (adopted)&$\Gamma_{\gamma}$ (meV) & Ref. &$\Gamma_{\alpha}$ (meV)\\
\hline
4.033 &$\frac{3}{2}^+$&$< 4.3 \times10^{-4}$& & &$< 4.3 \times10^{-4}$&45 $^{+200}_{-33}$& \cite{hackman00}&$< 0.13$\\
4.379 &$\frac{7}{2}^+$&$< 3.9 \times10^{-3}$&0.044 $\pm$ 0.032 & &$< 3.9 \times10^{-3}$&458 $\pm$ 92 & \cite{brown02}&$< 2.3$\\
4.549 &($\frac{3}{2})^-$&  0.16 $\pm$ 0.04 & 0.07 $\pm$ 0.03 & &0.10 $\pm$ 0.02 &39 $^{+34}_{-15}$ & \cite{tilley95}&4.4 $^{+ 4.0}_{- 2.0}$\\
4.600 &($\frac{5}{2}^+$)& 0.32 $\pm$ 0.04& 0.25 $\pm$ 0.04 & $0.32\pm0.03$ & 0.30 $\pm$ 0.02&101 $\pm$ 55 & \cite{kiss82} &43 $\pm$ 24\\
4.712 &($\frac{5}{2}^-$)& 0.85 $\pm$ 0.04 & 0.82 $\pm$ 0.15 & & 0.85 $\pm$ 0.04&43 $\pm$ 8& \cite{tilley95}&230 $\pm$ 80\\\
5.092 &$\frac{5}{2}^+$& 0.90 $\pm$ 0.06 & 0.90 $\pm$ 0.09 & & 0.90 $\pm$ 0.05&107 $\pm$ 17 & \cite{wilmes02,pringle89} &960 $\pm$ 530\\
\end{tabular}
\end{ruledtabular}
\end{table*}

\section{Decay Widths}

Very few experimental data on the lifetimes or radiative widths of states in $^{19}$Ne have been reported. The 4.033 MeV state is the only one of the six states considered here for which direct experimental measurements  are available. The radiative width $\Gamma_{\gamma}$ reported for this state \cite{hackman00} results from a combined analysis of Coulomb excitation and Doppler shift attenuation data \cite{davidson73}. In Ref.\ \cite{hackman00}, these experimental data are combined with shell-model calculations of the $E2/M1$ mixing ratio $\delta$ in order to reduce the uncertainties on the best-fit value. Since it is difficult to evaluate the theoretical uncertainties associated with this procedure, we adopt here the result of the analysis that assumes $\delta$ to be completely unknown and is therefore based entirely on experimental data, $\Gamma_{\gamma} = 45^{+200}_{-33}$ meV \cite{hackman00}. Note that this choice differs from that made in Ref.\ \cite{davids03}. We wish to be as conservative as possible here, so we opt for the value of $\Gamma_{\gamma}$ based purely on experimental measurements, without theoretical input and its corresponding uncertainty.

Measurements of $\Gamma_{\gamma}$ for analog states in the mirror nucleus $^{19}$F can be found for four of the states \cite{tilley95,kiss82,wilmes02,pringle89}, and these have been adopted under the assumption that $\Gamma_{\gamma}(^{19}$Ne) = $\Gamma_{\gamma}(^{19}$F). In the case of the 5.092 MeV state, $\Gamma_{\gamma}(^{19}$F) has been derived from the measured resonance strength \cite{wilmes02} and the measured $\gamma$-decay branching ratio \cite{pringle89}. For the 4.379 MeV state, no measurements are available in either nucleus, and we adopt the results of shell-model calculations \cite{brown02}, assigning a 1$\sigma$ uncertainty of 20\% to the calculated electromagnetic transition rate. Table \ref{table1} shows the adopted values of $\Gamma_{\gamma}$ and $\Gamma_{\alpha}$, which has been calculated as $\Gamma_{\alpha} = [B_{\alpha}/(1-B_{\alpha})]\Gamma_{\gamma}$. The table contains central values and 1$\sigma$ uncertainties where they can be calculated, and upper limits where this is not possible. These 90\% confidence level upper limits on $\Gamma_{\alpha}$ are based on 1.28$\sigma$ upper limits on $\Gamma_{\gamma}$. For comparison, Refs.\ \cite{wilmes95,oliveira97} contain earlier compilations of decay widths.

\section{Isospin Symmetry}

It is of interest to compare the decay widths of states in $^{19}$Ne with the corresponding widths in the mirror nucleus $^{19}$F. This issue has been addressed in the past \cite{oliveira97}, but the availability of new data on $^{19}$F \cite{wilmes02} makes it opportune to reexamine this question. As no experimental electromagnetic transition strengths for both a $^{19}$Ne state and its analog in $^{19}$F are available, we instead use all available information to compare their reduced alpha widths $\theta^2_{\alpha}$. We compute these reduced widths via  \begin{equation}
\label{ eqn1}
\theta^2_{\alpha}=\frac{\Gamma_{\ell}(E) R_n}{2 \hbar P_{\ell}(E, R_{n})} \sqrt{\frac{\mu}{2 E}}, 
\end{equation} where $\Gamma_{\ell}$(E) is the energy- and angular momentum-dependent partial width, R$_n$ the nuclear radius, P$_{\ell}$(E, R$_n$) the Coulomb penetrability, $\mu$ the reduced mass, and E the center of mass energy of the resonance \cite{rolfs88}. The penetrability is given by \begin{equation}
\label{eqn2 }
P_{\ell}(E, R_{n}) = \frac{1}{F^{2}_{l}(E, R_n) + G^{2}_{l}(E, R_n)}, 
\end{equation} where $F$ and $G$ are the regular and irregular Coulomb wavefunctions respectively. For simplicity, we assume that R$_n$ = 5 fm for both $^{19}$Ne and $^{19}$F, as was done in Ref.\ \cite{oliveira97}. The results of this comparison are shown in Table \ref{table2}. For all of the analog states in the T~=~1/2, A~=~19 system examined here, the reduced $\alpha$ widths agree within the experimental uncertainties at the 2$\sigma$ level.

\begin{table*}
 \caption{\label{table2}Decay and reduced $\alpha$ widths in $^{19}$Ne and $^{19}$F. Upper limits are specified at the 90\% confidence level.}
 \begin{ruledtabular}
 \begin{tabular}{ccccccc}
$^{19}$Ne E$_{x}$ (MeV) & $^{19}$F E$_{x}$ (MeV) & J$^{\pi}$ & $^{19}$Ne $\Gamma_{\alpha}$ (meV) & $^{19}$F $\Gamma_{\alpha}$ (meV) (\cite{wilmes02}) & $\theta^2_{\alpha}$ ($^{19}$Ne) & $\theta^2_{\alpha}$ ($^{19}$F)\\
\hline
4.379 & 4.378 & $\frac{7}{2}^+$ & $< 2.3$ & $1.5^{+1.5}_{-0.8}\times10^{-6}$ & $<0.095$ & 0.0078$^{+0.0078}_{-0.004}$\\
4.549 & 4.556 & ($\frac{3}{2})^-$ & 4.4$^{+ 4.0}_{- 2.0}$ & 3.2$\pm1.3\times10^{-3}$ & 0.0016$^{+0.0015}_{-0.0007}$ & $0.0009\pm0.0004$\\
4.600 & 4.550 & ($\frac{5}{2}^+$) & $43\pm24$ & $32\pm4\times10^{-3}$ & $0.063\pm0.035$ & $0.12\pm0.01$\\
4.712 & 4.683 & ($\frac{5}{2}^-$) & $230\pm80$ & $1.9\pm0.2$ & $0.012\pm0.004$ & $0.020\pm0.002$\\
5.092 & 5.107 &$ \frac{5}{2}^+$ & $960\pm530$ & $3.3\pm0.6$ & $0.013\pm0.007$ & $0.00048\pm0.00009$\\
\end{tabular}
\end{ruledtabular}
\end{table*}
 
 \section{Astrophysical Implications}

\begin{figure}\includegraphics[width=\linewidth]{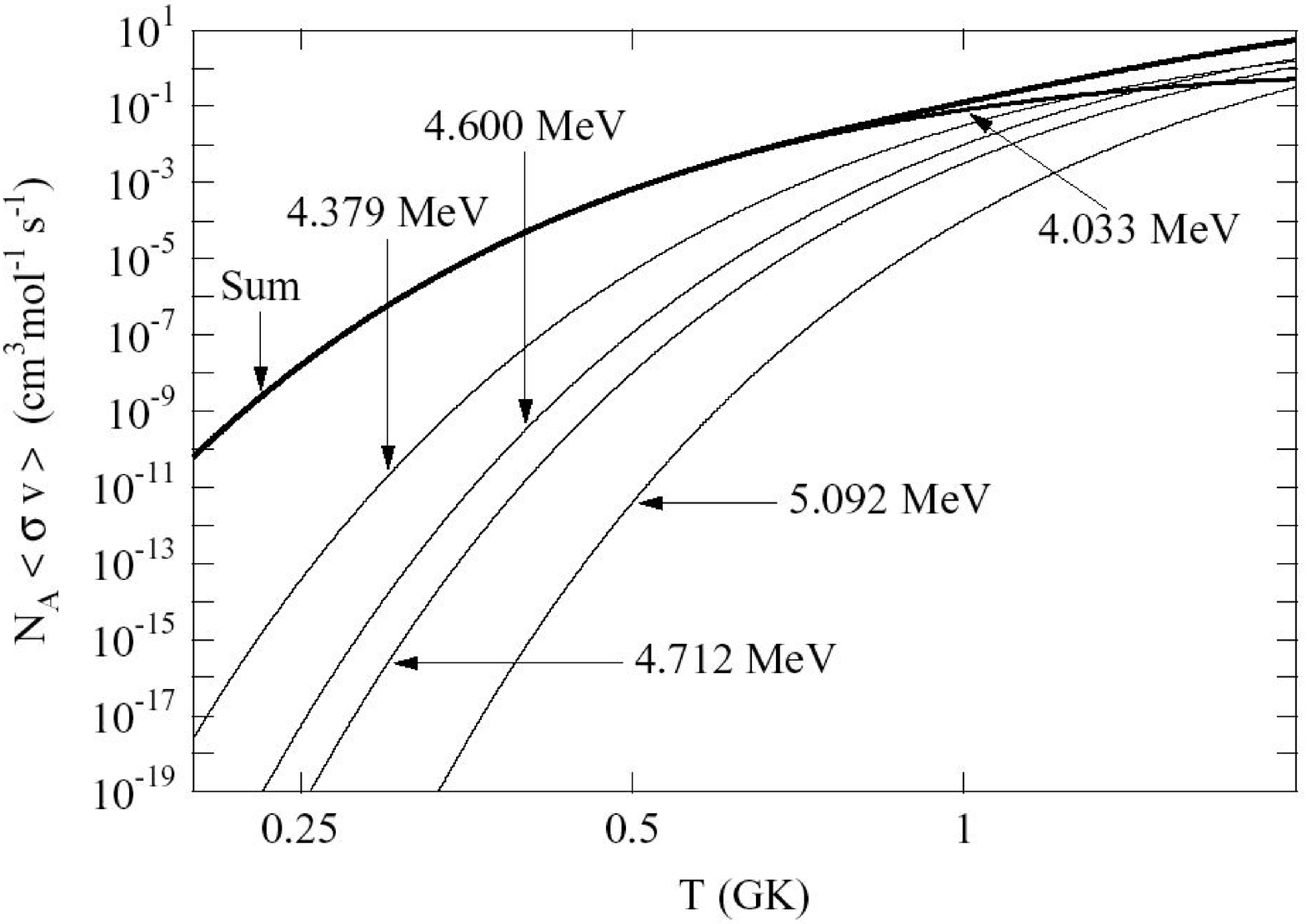} \caption{Product of the Avogadro constant N$_{\mathrm{A}}$ and the thermally averaged rate of the $^{15}$O($\alpha$,$\gamma$)$^{19}$Ne reaction per particle pair. Contributions from all of the important states are shown, along with the sum of the resonant and direct capture rates. The contributions of the 4.033 and 4.379 MeV states are 99.73\% confidence level upper limits calculated as described in the text.} \label{fig10} \end{figure}

We have calculated the thermally averaged $^{15}$O($\alpha,\gamma)^{19}$Ne reaction rate per particle pair as \begin{equation}
\label{ eqn3}<\sigma v>=(\frac{2 \pi}{\mu kT})^{\frac{3}{2}} \hbar^{2} \sum_{i} (\omega\gamma)_{i} \mathrm{exp}(\frac{-E_{i}}{kT}),\end{equation} where the sum runs over the resonances whose strengths are given by \begin{equation}
\label{eqn4 }
\omega\gamma=\frac{2J +1}{(2J_{^{15}O}+1)(2J_{\alpha}+1)}\frac{\Gamma_{\alpha}\Gamma_{\gamma}}{\Gamma_{\alpha}+\Gamma_{\gamma}}.\end{equation} In these equations, $\sigma$ is the reaction cross section, $v$ the relative velocity, $k$ the Boltzmann constant, $T$ the temperature, $E$ the center of mass energy of a resonance and $J$ its angular momentum. The contributions of individual states are shown in Fig.\ \ref{fig10}. Since we wish to be as conservative as possible in evaluating the reaction rate, the contributions of the 4.033- and 4.379 MeV states are calculated using our 99.73\% confidence level upper limits on their $\alpha$ widths, 514 $\mu$eV and 5.6 meV respectively. These values were obtained using our upper limits on B$_{\alpha}$ and 3$\sigma$ upper limits on $\Gamma_{\gamma}$ for these states. Also shown in Fig.\ \ref{fig10} is the sum of the individual rates and the direct capture contribution, which was calculated as per the prescription of Ref.\ \cite{langanke86}. Although the contribution of the 4.549 MeV state is included in the sum, its individual contribution is not shown in the figure because it is so small. The 4.033 MeV state completely dominates the reaction rate at temperatures below 0.5 GK. Since hydrodynamic nova models indicate that peak temperatures do not exceed this value \cite{jose98,starrfield00}, the 4.033 MeV state is the only resonance that need be considered in calculating the reaction rate in novae.

The possibility of breakout from the hot CNO cycles into the $rp$ process under the high temperature and density conditions attained in the most massive ONe nova outbursts has been extensively discussed in previous work \cite{wiescher99}. In order to investigate this possibility, we have performed a series of hydrodynamic calculations of nova outbursts on 1.35 M$_\odot$ ONe white dwarfs using an implicit, spherically symmetric, hydrodynamic code in Lagrangian formulation \cite{jose98}. This code describes the entire outburst from the onset of accretion through the thermonuclear runaway, including the expansion and ejection of the accreted envelope. We assume a mass accretion rate of $2\times10^{-10}$ M$_\odot$ yr$^{-1}$ and 50\% mixing between the solar-like accreted material and the outermost, ONe-rich shells of the underlying white dwarf.  The mass, composition, and mass accretion rate were chosen in order to simulate the most energetic, hottest novae, in which hot CNO breakout is the most likely. In the same spirit, we have employed in these simulations an extreme, 99.73\% confidence level upper limit on the $^{15}$O($\alpha$,$\gamma$)$^{19}$Ne rate. This rate is a factor of $\sim50$ larger than that adopted in previous work \cite{langanke86,mao95,vancraeynest98}. For these reasons, our simulations would identify the role played by $^{15}$O($\alpha$,$\gamma$)$^{19}$Ne in hot CNO breakout in novae, if there were any.

\begin{figure*}\includegraphics[width=\linewidth]{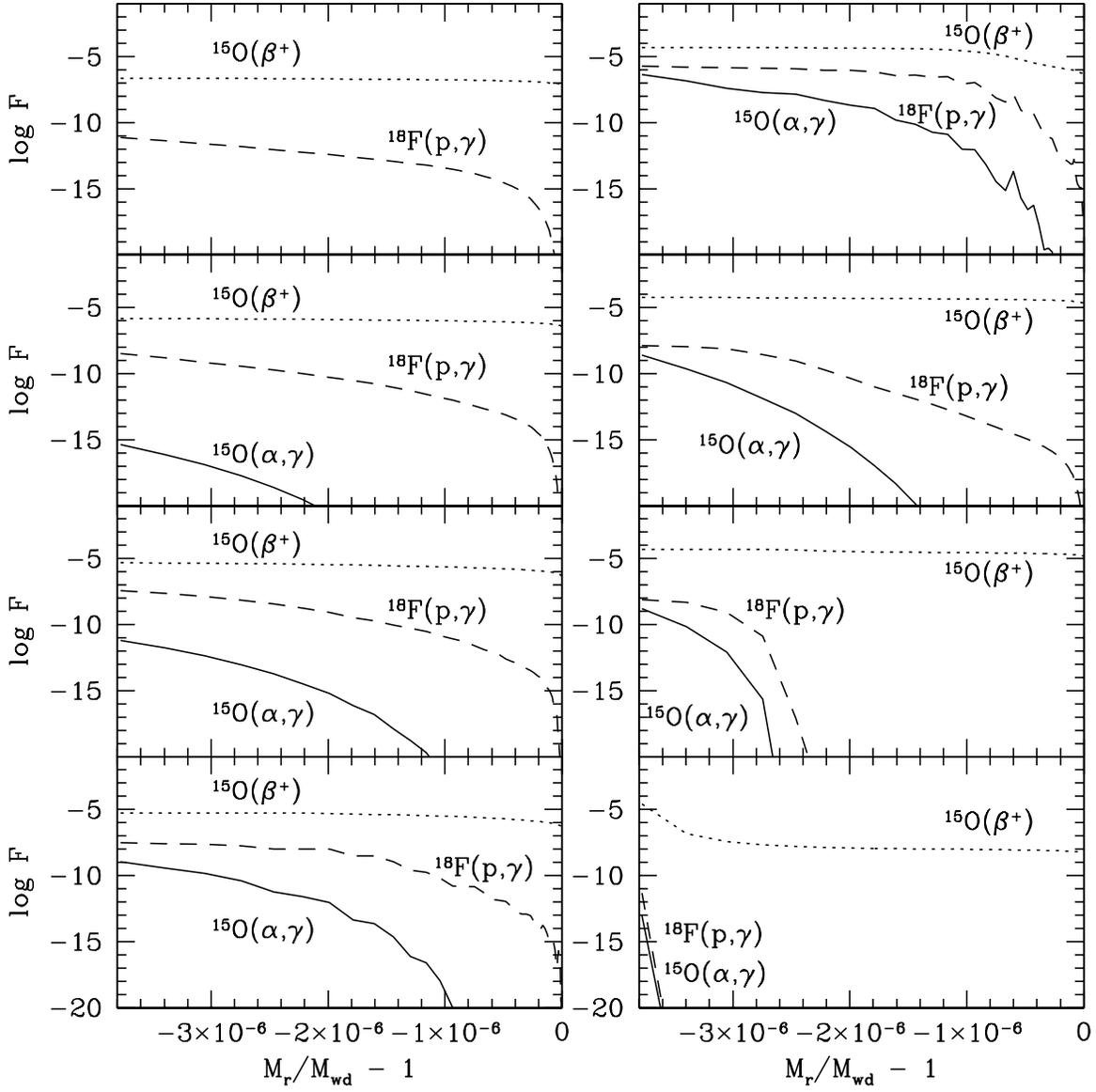} \caption{Reaction and decay rates per unit volume (cm$^{-3}$ s$^{-1}$) in logarithmic scale along the accreted envelope for a 1.35 solar mass ONe nova accreting at a rate $\dot M = 2 \times 10^{-10}$ $M_\odot$ yr$^{-1}$. The mass coordinate represents the mass below the surface (M$_r$) relative to the total white dwarf mass (M$_{wd}$). Thus the left side of each panel corresponds to the base of the accreted envelope, and the right side its outer surface. From top to bottom and then left to right, the first five panels show a time series from the early stages of the explosion up to the ejection stage, with a temperature in the burning shell equal to: 0.1, 0.15, 0.2, 0.25, and $T_{peak} = 0.32$ GK. The last three panels correspond to the final phases of the explosion, when the white dwarf envelope has expanded to a size of $R_{wd} \sim 10^9, 10^{10}$, and $10^{12}$ cm, respectively. All of the material lying at a mass coordinate $M_r/M_{wd}-1\geq-3.05\times10^{-6}$ is ejected by the explosion.} \label{fig11} \end{figure*}

In order to evaluate the potential of the $^{15}$O($\alpha$,$\gamma$) reaction to initiate breakout from the CNO cycles, we compare its rate with the $^{15}$O $\beta^+$ decay rate. We compute the rate per unit volume of a reaction involving distinct nuclei 1 and 2 \begin{equation}\label{eqn5}F=n_1n_2 <\sigma v>=\frac{\rho N_{\mathrm{A}} X_1}{A_1}\frac{\rho N_{\mathrm{A}} X_2}{A_2}<\sigma v>,\end{equation} where $n_i$ is the number density, $X_i$ the mass fraction, and $A_i$ the atomic mass of nucleus i, and $\rho$ is the mass density. The $\beta^+$ decay rate of a nucleus per unit volume  is given by\begin{equation}
\label{ eqn6}
F=n\lambda=\frac{\rho N_{\mathrm{A}} X}{A}\frac{\mathrm{ln}(2)}{t_{1/2}},\end{equation} where $t_{1/2}$ is its $\beta^+$ decay half-life. As shown in Fig.\ \ref{fig11}, the $^{15}$O($\alpha$,$\gamma$)$^{19}$Ne rate is less than 0.3\% of the $^{15}$O $\beta^+$ decay rate at all times throughout the whole accreted envelope, including the hottest, innermost layer at its base, which is not ejected by the explosion. Hence more than 99.7\% of the $^{15}$O nuclei synthesized in the nova outburst $\beta^+$ decay before they capture an $\alpha$ particle, and there is no significant CNO breakout via $^{15}$O($\alpha$,$\gamma$)$^{19}$Ne. Moreover, as Fig.\ \ref{fig11} illustrates, the contribution of $^{15}$O($\alpha$,$\gamma$) to the production of $^{19}$Ne is greatly exceeded by the contribution of $^{18}$F($p,\gamma$), the rate of which is taken from Ref.\ \cite{coc00}. Thus our nova simulations support the conclusion drawn in Refs.\ \cite{wiescher99,davids03,iliadis02} that the $^{15}$O($\alpha,\gamma$)$^{19}$Ne reaction plays no important role in nova explosions.
 
\section{Summary}

In summary, we have measured the $\alpha$-decay branching ratios for all of the states in $^{19}$Ne relevant to the astrophysical rate of the $^{15}$O($\alpha$,$\gamma$)$^{19}$Ne reaction, populating them by means of the $p(^{21}$Ne,$t)^{19}$Ne reaction at 43 MeV/u and observing their decays with 100\% geometric efficiency in a magnetic spectrometer. Combining our measurements with independent determinations of the radiative widths of these states, we have  calculated the astrophysical rate of $^{15}$O($\alpha$,$\gamma$)$^{19}$Ne. Using this rate, we have carried out a series of hydrodynamic simulations of nova outbursts. On the basis of these calculations, we conclude that there can be no significant enrichment of nova ejecta due to CNO breakout via $^{15}$O($\alpha$,$\gamma$)$^{19}$Ne.

\begin{acknowledgments}
This work was performed as part of the research program of the {\it Stichting voor Fundamenteel Onderzoek der Materie} with financial support from the {\it Nederlandse Organisatie voor Wetenschappelijk Onderzoek}. KER, RHS, and AHW acknowledge support from a NATO Collaborative Linkage Grant. The ANL Physics Division is supported by the U. S. Department of Energy Nuclear Physics Division under Contract No. W-31-109-ENG-38. RES acknowledges support from the US DOE under contract DE-FG02-98ER41086. JJ and MH acknowledge support from MCYT grants AYA2001-2360, AYA2002-04094-C03-02 and AYA2002-04094-C03-03.
\end{acknowledgments}

\bibliography{s23}

\begin{thebibliography}{34}
\expandafter\ifx\csname natexlab\endcsname\relax\def\natexlab#1{#1}\fi
\expandafter\ifx\csname bibnamefont\endcsname\relax
  \def\bibnamefont#1{#1}\fi
\expandafter\ifx\csname bibfnamefont\endcsname\relax
  \def\bibfnamefont#1{#1}\fi
\expandafter\ifx\csname citenamefont\endcsname\relax
  \def\citenamefont#1{#1}\fi
\expandafter\ifx\csname url\endcsname\relax
  \def\url#1{\texttt{#1}}\fi
\expandafter\ifx\csname urlprefix\endcsname\relax\def\urlprefix{URL }\fi
\providecommand{\bibinfo}[2]{#2}
\providecommand{\eprint}[2][]{\url{#2}}

\bibitem[{\citenamefont{Jos\'{e} and Hernanz}(1998)}]{jose98}
\bibinfo{author}{\bibfnamefont{J.}~\bibnamefont{Jos\'{e}}} \bibnamefont{and}
  \bibinfo{author}{\bibfnamefont{M.}~\bibnamefont{Hernanz}},
  \bibinfo{journal}{Astrophys. J.} \textbf{\bibinfo{volume}{494}},
  \bibinfo{pages}{680} (\bibinfo{year}{1998}).

\bibitem[{\citenamefont{{Starrfield} et~al.}(2000)\citenamefont{{Starrfield},
  {Sparks}, {Truran}, and {Wiescher}}}]{starrfield00}
\bibinfo{author}{\bibfnamefont{S.}~\bibnamefont{{Starrfield}}},
  \bibinfo{author}{\bibfnamefont{W.~M.} \bibnamefont{{Sparks}}},
  \bibinfo{author}{\bibfnamefont{J.~W.} \bibnamefont{{Truran}}},
  \bibnamefont{and} \bibinfo{author}{\bibfnamefont{M.~C.}
  \bibnamefont{{Wiescher}}}, \bibinfo{journal}{Astrophys. J. Suppl. Ser.}
  \textbf{\bibinfo{volume}{127}}, \bibinfo{pages}{485} (\bibinfo{year}{2000}).

\bibitem[{\citenamefont{{Jos{\' e}} et~al.}(2001)\citenamefont{{Jos{\' e}},
  {Coc}, and {Hernanz}}}]{jose01}
\bibinfo{author}{\bibfnamefont{J.}~\bibnamefont{{Jos{\' e}}}},
  \bibinfo{author}{\bibfnamefont{A.}~\bibnamefont{{Coc}}}, \bibnamefont{and}
  \bibinfo{author}{\bibfnamefont{M.}~\bibnamefont{{Hernanz}}},
  \bibinfo{journal}{\apj} \textbf{\bibinfo{volume}{560}}, \bibinfo{pages}{897}
  (\bibinfo{year}{2001}).

\bibitem[{\citenamefont{Wallace and Woosley}(1981)}]{wallace81}
\bibinfo{author}{\bibfnamefont{R.~K.} \bibnamefont{Wallace}} \bibnamefont{and}
  \bibinfo{author}{\bibfnamefont{S.~E.} \bibnamefont{Woosley}},
  \bibinfo{journal}{Astrophys. J. Suppl. Ser.} \textbf{\bibinfo{volume}{45}},
  \bibinfo{pages}{389} (\bibinfo{year}{1981}).

\bibitem[{\citenamefont{{Schatz} et~al.}(1998)\citenamefont{{Schatz},
  {Aprahamian}, {G\"{o}rres}, {Wiescher}, {Rauscher}, {Rembges}, {Thielemann},
  {Pfeiffer}, {Moeller}, {Kratz} et~al.}}]{schatz98}
\bibinfo{author}{\bibfnamefont{H.}~\bibnamefont{{Schatz}}},
  \bibinfo{author}{\bibfnamefont{A.}~\bibnamefont{{Aprahamian}}},
  \bibinfo{author}{\bibfnamefont{J.}~\bibnamefont{{G\"{o}rres}}},
  \bibinfo{author}{\bibfnamefont{M.}~\bibnamefont{{Wiescher}}},
  \bibinfo{author}{\bibfnamefont{T.}~\bibnamefont{{Rauscher}}},
  \bibinfo{author}{\bibfnamefont{J.~F.} \bibnamefont{{Rembges}}},
  \bibinfo{author}{\bibfnamefont{F.-K.} \bibnamefont{{Thielemann}}},
  \bibinfo{author}{\bibfnamefont{B.}~\bibnamefont{{Pfeiffer}}},
  \bibinfo{author}{\bibfnamefont{P.}~\bibnamefont{{Moeller}}},
  \bibinfo{author}{\bibfnamefont{K.-L.} \bibnamefont{{Kratz}}},
  \bibnamefont{et~al.}, \bibinfo{journal}{Phys. Rep.}
  \textbf{\bibinfo{volume}{294}}, \bibinfo{pages}{167} (\bibinfo{year}{1998}).

\bibitem[{\citenamefont{{Schatz} et~al.}(2001)\citenamefont{{Schatz},
  {Aprahamian}, {Barnard}, {Bildsten}, {Cumming}, {Ouellette}, {Rauscher},
  {Thielemann}, and {Wiescher}}}]{schatz01}
\bibinfo{author}{\bibfnamefont{H.}~\bibnamefont{{Schatz}}},
  \bibinfo{author}{\bibfnamefont{A.}~\bibnamefont{{Aprahamian}}},
  \bibinfo{author}{\bibfnamefont{V.}~\bibnamefont{{Barnard}}},
  \bibinfo{author}{\bibfnamefont{L.}~\bibnamefont{{Bildsten}}},
  \bibinfo{author}{\bibfnamefont{A.}~\bibnamefont{{Cumming}}},
  \bibinfo{author}{\bibfnamefont{M.}~\bibnamefont{{Ouellette}}},
  \bibinfo{author}{\bibfnamefont{T.}~\bibnamefont{{Rauscher}}},
  \bibinfo{author}{\bibfnamefont{F.-K.} \bibnamefont{{Thielemann}}},
  \bibnamefont{and}
  \bibinfo{author}{\bibfnamefont{M.}~\bibnamefont{{Wiescher}}},
  \bibinfo{journal}{Phys. Rev. Lett.} \textbf{\bibinfo{volume}{86}},
  \bibinfo{pages}{3471} (\bibinfo{year}{2001}).

\bibitem[{\citenamefont{Wiescher et~al.}(1999)\citenamefont{Wiescher,
  G\"{o}rres, and Schatz}}]{wiescher99}
\bibinfo{author}{\bibfnamefont{M.}~\bibnamefont{Wiescher}},
  \bibinfo{author}{\bibfnamefont{J.}~\bibnamefont{G\"{o}rres}},
  \bibnamefont{and} \bibinfo{author}{\bibfnamefont{H.}~\bibnamefont{Schatz}},
  \bibinfo{journal}{J. Phys. G} \textbf{\bibinfo{volume}{25}},
  \bibinfo{pages}{R133} (\bibinfo{year}{1999}).

\bibitem[{\citenamefont{Audi and Wapstra}(1995)}]{audi95}
\bibinfo{author}{\bibfnamefont{G.}~\bibnamefont{Audi}} \bibnamefont{and}
  \bibinfo{author}{\bibfnamefont{A.~H.} \bibnamefont{Wapstra}},
  \bibinfo{journal}{Nucl. Phys.} \textbf{\bibinfo{volume}{A595}},
  \bibinfo{pages}{409} (\bibinfo{year}{1995}).

\bibitem[{\citenamefont{Magnus et~al.}(1990)\citenamefont{Magnus, Smith,
  Howard, Parker, and Champagne}}]{magnus90}
\bibinfo{author}{\bibfnamefont{P.~V.} \bibnamefont{Magnus}},
  \bibinfo{author}{\bibfnamefont{M.~S.} \bibnamefont{Smith}},
  \bibinfo{author}{\bibfnamefont{A.~J.} \bibnamefont{Howard}},
  \bibinfo{author}{\bibfnamefont{P.~D.} \bibnamefont{Parker}},
  \bibnamefont{and} \bibinfo{author}{\bibfnamefont{A.~E.}
  \bibnamefont{Champagne}}, \bibinfo{journal}{Nucl. Phys.}
  \textbf{\bibinfo{volume}{A506}}, \bibinfo{pages}{332} (\bibinfo{year}{1990}).

\bibitem[{\citenamefont{{Kubono} et~al.}(2002)\citenamefont{{Kubono},
  {Yanagisawa}, {Teranishi}, {Kato}, {Kishida}, {Michimasa}, {Ohshiro},
  {Shimoura}, {Ue}, {Watanabe} et~al.}}]{kubono02}
\bibinfo{author}{\bibfnamefont{S.}~\bibnamefont{{Kubono}}},
  \bibinfo{author}{\bibfnamefont{Y.}~\bibnamefont{{Yanagisawa}}},
  \bibinfo{author}{\bibfnamefont{T.}~\bibnamefont{{Teranishi}}},
  \bibinfo{author}{\bibfnamefont{S.}~\bibnamefont{{Kato}}},
  \bibinfo{author}{\bibfnamefont{Y.}~\bibnamefont{{Kishida}}},
  \bibinfo{author}{\bibfnamefont{S.}~\bibnamefont{{Michimasa}}},
  \bibinfo{author}{\bibfnamefont{Y.}~\bibnamefont{{Ohshiro}}},
  \bibinfo{author}{\bibfnamefont{S.}~\bibnamefont{{Shimoura}}},
  \bibinfo{author}{\bibfnamefont{K.}~\bibnamefont{{Ue}}},
  \bibinfo{author}{\bibfnamefont{S.}~\bibnamefont{{Watanabe}}},
  \bibnamefont{et~al.}, \bibinfo{journal}{European Physical Journal A}
  \textbf{\bibinfo{volume}{13}}, \bibinfo{pages}{217} (\bibinfo{year}{2002}).

\bibitem[{\citenamefont{{Laird} et~al.}(2002)\citenamefont{{Laird},
  {Cherubini}, {Ostrowski}, {Aliotta}, {Davinson}, {di Pietro}, {Figuera},
  {Galster}, {Graulich}, {Groombridge} et~al.}}]{laird02}
\bibinfo{author}{\bibfnamefont{A.~M.} \bibnamefont{{Laird}}},
  \bibinfo{author}{\bibfnamefont{S.}~\bibnamefont{{Cherubini}}},
  \bibinfo{author}{\bibfnamefont{A.~N.} \bibnamefont{{Ostrowski}}},
  \bibinfo{author}{\bibfnamefont{M.}~\bibnamefont{{Aliotta}}},
  \bibinfo{author}{\bibfnamefont{T.}~\bibnamefont{{Davinson}}},
  \bibinfo{author}{\bibfnamefont{A.}~\bibnamefont{{di Pietro}}},
  \bibinfo{author}{\bibfnamefont{P.}~\bibnamefont{{Figuera}}},
  \bibinfo{author}{\bibfnamefont{W.}~\bibnamefont{{Galster}}},
  \bibinfo{author}{\bibfnamefont{J.~S.} \bibnamefont{{Graulich}}},
  \bibinfo{author}{\bibfnamefont{D.}~\bibnamefont{{Groombridge}}},
  \bibnamefont{et~al.}, \bibinfo{journal}{Phys. Rev. C}
  \textbf{\bibinfo{volume}{66}}, \bibinfo{pages}{048801}
  (\bibinfo{year}{2002}).

\bibitem[{\citenamefont{Mao et~al.}(1995)\citenamefont{Mao, Fortune, and
  Lacaze}}]{mao95}
\bibinfo{author}{\bibfnamefont{Z.~Q.} \bibnamefont{Mao}},
  \bibinfo{author}{\bibfnamefont{H.~T.} \bibnamefont{Fortune}},
  \bibnamefont{and} \bibinfo{author}{\bibfnamefont{A.~G.}
  \bibnamefont{Lacaze}}, \bibinfo{journal}{Phys. Rev. Lett.}
  \textbf{\bibinfo{volume}{74}}, \bibinfo{pages}{3760} (\bibinfo{year}{1995}).

\bibitem[{\citenamefont{Mao et~al.}(1996)\citenamefont{Mao, Fortune, and
  Lacaze}}]{mao96}
\bibinfo{author}{\bibfnamefont{Z.~Q.} \bibnamefont{Mao}},
  \bibinfo{author}{\bibfnamefont{H.~T.} \bibnamefont{Fortune}},
  \bibnamefont{and} \bibinfo{author}{\bibfnamefont{A.~G.}
  \bibnamefont{Lacaze}}, \bibinfo{journal}{Phys. Rev. C}
  \textbf{\bibinfo{volume}{53}}, \bibinfo{pages}{1197} (\bibinfo{year}{1996}).

\bibitem[{\citenamefont{Davids et~al.}(2003)\citenamefont{Davids, van~den Berg,
  Dendooven, Fleurot, Hunyadi, de~Huu, Rehm, Segel, Siemssen, Wilschut
  et~al.}}]{davids03}
\bibinfo{author}{\bibfnamefont{B.}~\bibnamefont{Davids}},
  \bibinfo{author}{\bibfnamefont{A.~M.} \bibnamefont{van~den Berg}},
  \bibinfo{author}{\bibfnamefont{P.}~\bibnamefont{Dendooven}},
  \bibinfo{author}{\bibfnamefont{F.}~\bibnamefont{Fleurot}},
  \bibinfo{author}{\bibfnamefont{M.}~\bibnamefont{Hunyadi}},
  \bibinfo{author}{\bibfnamefont{M.~A.} \bibnamefont{de~Huu}},
  \bibinfo{author}{\bibfnamefont{K.~E.} \bibnamefont{Rehm}},
  \bibinfo{author}{\bibfnamefont{R.~E.} \bibnamefont{Segel}},
  \bibinfo{author}{\bibfnamefont{R.~H.} \bibnamefont{Siemssen}},
  \bibinfo{author}{\bibfnamefont{H.~W.} \bibnamefont{Wilschut}},
  \bibnamefont{et~al.}, \bibinfo{journal}{\prc} \textbf{\bibinfo{volume}{67}},
  \bibinfo{pages}{012801(R)} (\bibinfo{year}{2003}).

\bibitem[{\citenamefont{Rehm et~al.}(2000)\citenamefont{Rehm, Caggiano, Collon,
  Heinz, Janssens, Jiang, Pardo, Paul, Schiffer, Siemssen et~al.}}]{rehm00}
\bibinfo{author}{\bibfnamefont{K.~E.} \bibnamefont{Rehm}},
  \bibinfo{author}{\bibfnamefont{J.~A.} \bibnamefont{Caggiano}},
  \bibinfo{author}{\bibfnamefont{P.}~\bibnamefont{Collon}},
  \bibinfo{author}{\bibfnamefont{A.}~\bibnamefont{Heinz}},
  \bibinfo{author}{\bibfnamefont{R.~V.~F.} \bibnamefont{Janssens}},
  \bibinfo{author}{\bibfnamefont{C.~L.} \bibnamefont{Jiang}},
  \bibinfo{author}{\bibfnamefont{R.}~\bibnamefont{Pardo}},
  \bibinfo{author}{\bibfnamefont{M.}~\bibnamefont{Paul}},
  \bibinfo{author}{\bibfnamefont{J.~P.} \bibnamefont{Schiffer}},
  \bibinfo{author}{\bibfnamefont{R.~H.} \bibnamefont{Siemssen}},
  \bibnamefont{et~al.}, \emph{\bibinfo{title}{Argonne National Laboratory
  Physics Division Annual Report}} (\bibinfo{publisher}{unpublished},
  \bibinfo{year}{2000}), p.~\bibinfo{pages}{6}.

\bibitem[{\citenamefont{van~den Berg}(1995)}]{berg95}
\bibinfo{author}{\bibfnamefont{A.~M.} \bibnamefont{van~den Berg}},
  \bibinfo{journal}{Nucl. Instrum. Methods} \textbf{\bibinfo{volume}{B99}},
  \bibinfo{pages}{637} (\bibinfo{year}{1995}).

\bibitem[{\citenamefont{Leegte et~al.}(1992)\citenamefont{Leegte, Koldenhof,
  Boonstra, and Wilschut}}]{leegte92}
\bibinfo{author}{\bibfnamefont{H.~K.~W.} \bibnamefont{Leegte}},
  \bibinfo{author}{\bibfnamefont{E.~E.} \bibnamefont{Koldenhof}},
  \bibinfo{author}{\bibfnamefont{A.~L.} \bibnamefont{Boonstra}},
  \bibnamefont{and} \bibinfo{author}{\bibfnamefont{H.~W.}
  \bibnamefont{Wilschut}}, \bibinfo{journal}{Nucl. Instrum. Methods}
  \textbf{\bibinfo{volume}{A313}}, \bibinfo{pages}{260} (\bibinfo{year}{1992}).

\bibitem[{\citenamefont{W\"{o}rtche}(2001)}]{woertche01}
\bibinfo{author}{\bibfnamefont{H.~J.} \bibnamefont{W\"{o}rtche}},
  \bibinfo{journal}{Nucl. Phys.} \textbf{\bibinfo{volume}{A687}},
  \bibinfo{pages}{321c} (\bibinfo{year}{2001}).

\bibitem[{\citenamefont{{Tilley} et~al.}(1995)\citenamefont{{Tilley}, {Weller},
  {Cheves}, and {Chasteler}}}]{tilley95}
\bibinfo{author}{\bibfnamefont{D.~R.} \bibnamefont{{Tilley}}},
  \bibinfo{author}{\bibfnamefont{H.~R.} \bibnamefont{{Weller}}},
  \bibinfo{author}{\bibfnamefont{C.~M.} \bibnamefont{{Cheves}}},
  \bibnamefont{and} \bibinfo{author}{\bibfnamefont{R.~M.}
  \bibnamefont{{Chasteler}}}, \bibinfo{journal}{Nucl. Phys.}
  \textbf{\bibinfo{volume}{A595}}, \bibinfo{pages}{1} (\bibinfo{year}{1995}).

\bibitem[{\citenamefont{{Fortune} et~al.}(1978)\citenamefont{{Fortune}, {Nann},
  and {Wildenthal}}}]{fortune78}
\bibinfo{author}{\bibfnamefont{H.~T.} \bibnamefont{{Fortune}}},
  \bibinfo{author}{\bibfnamefont{H.}~\bibnamefont{{Nann}}}, \bibnamefont{and}
  \bibinfo{author}{\bibfnamefont{B.~H.} \bibnamefont{{Wildenthal}}},
  \bibinfo{journal}{\prc} \textbf{\bibinfo{volume}{18}}, \bibinfo{pages}{1563}
  (\bibinfo{year}{1978}).

\bibitem[{\citenamefont{{Hagiwara} et~al.}(2002)}]{hagiwara02}
\bibinfo{author}{\bibfnamefont{K.}~\bibnamefont{{Hagiwara}}}
  \bibnamefont{et~al.}, \bibinfo{journal}{\prd} \textbf{\bibinfo{volume}{66}},
  \bibinfo{pages}{010001} (\bibinfo{year}{2002}).

\bibitem[{\citenamefont{{Hackman} et~al.}(2000)\citenamefont{{Hackman},
  {Austin}, {Glasmacher}, {Aumann}, {Brown}, {Ibbotson}, {Miller},
  {Pritychenko}, {Riley}, {Roeder} et~al.}}]{hackman00}
\bibinfo{author}{\bibfnamefont{G.}~\bibnamefont{{Hackman}}},
  \bibinfo{author}{\bibfnamefont{S.~M.} \bibnamefont{{Austin}}},
  \bibinfo{author}{\bibfnamefont{T.}~\bibnamefont{{Glasmacher}}},
  \bibinfo{author}{\bibfnamefont{T.}~\bibnamefont{{Aumann}}},
  \bibinfo{author}{\bibfnamefont{B.~A.} \bibnamefont{{Brown}}},
  \bibinfo{author}{\bibfnamefont{R.~W.} \bibnamefont{{Ibbotson}}},
  \bibinfo{author}{\bibfnamefont{K.}~\bibnamefont{{Miller}}},
  \bibinfo{author}{\bibfnamefont{B.}~\bibnamefont{{Pritychenko}}},
  \bibinfo{author}{\bibfnamefont{L.~A.} \bibnamefont{{Riley}}},
  \bibinfo{author}{\bibfnamefont{B.}~\bibnamefont{{Roeder}}},
  \bibnamefont{et~al.}, \bibinfo{journal}{Phys. Rev. C}
  \textbf{\bibinfo{volume}{61}}, \bibinfo{pages}{052801(R)}
  (\bibinfo{year}{2000}).

\bibitem[{\citenamefont{Brown}(2002)}]{brown02}
\bibinfo{author}{\bibfnamefont{B.~A.} \bibnamefont{Brown}},
  \bibinfo{howpublished}{private communication} (\bibinfo{year}{2002}).

\bibitem[{\citenamefont{Kiss et~al.}(1982)\citenamefont{Kiss, Nyak\'o,
  Somorjai, Anttila, and Bister}}]{kiss82}
\bibinfo{author}{\bibfnamefont{{\'{A}}.~Z.} \bibnamefont{Kiss}},
  \bibinfo{author}{\bibfnamefont{B.}~\bibnamefont{Nyak\'o}},
  \bibinfo{author}{\bibfnamefont{E.}~\bibnamefont{Somorjai}},
  \bibinfo{author}{\bibfnamefont{A.}~\bibnamefont{Anttila}}, \bibnamefont{and}
  \bibinfo{author}{\bibfnamefont{M.}~\bibnamefont{Bister}},
  \bibinfo{journal}{Nucl. Instrum. Methods} \textbf{\bibinfo{volume}{203}},
  \bibinfo{pages}{107} (\bibinfo{year}{1982}).

\bibitem[{\citenamefont{{Wilmes} et~al.}(2002)\citenamefont{{Wilmes}, {Wilmes},
  {Staudt}, {Mohr}, and {Hammer}}}]{wilmes02}
\bibinfo{author}{\bibfnamefont{S.}~\bibnamefont{{Wilmes}}},
  \bibinfo{author}{\bibfnamefont{V.}~\bibnamefont{{Wilmes}}},
  \bibinfo{author}{\bibfnamefont{G.}~\bibnamefont{{Staudt}}},
  \bibinfo{author}{\bibfnamefont{P.}~\bibnamefont{{Mohr}}}, \bibnamefont{and}
  \bibinfo{author}{\bibfnamefont{J.~W.} \bibnamefont{{Hammer}}},
  \bibinfo{journal}{\prc} \textbf{\bibinfo{volume}{66}}, \bibinfo{pages}{65802}
  (\bibinfo{year}{2002}).

\bibitem[{\citenamefont{Pringle and Vermeer}(1989)}]{pringle89}
\bibinfo{author}{\bibfnamefont{D.~M.} \bibnamefont{Pringle}} \bibnamefont{and}
  \bibinfo{author}{\bibfnamefont{W.~J.} \bibnamefont{Vermeer}},
  \bibinfo{journal}{Nucl. Phys.} \textbf{\bibinfo{volume}{A499}},
  \bibinfo{pages}{117} (\bibinfo{year}{1989}).

\bibitem[{\citenamefont{Davidson and Roush}(1973)}]{davidson73}
\bibinfo{author}{\bibfnamefont{J.~M.} \bibnamefont{Davidson}} \bibnamefont{and}
  \bibinfo{author}{\bibfnamefont{M.~L.} \bibnamefont{Roush}},
  \bibinfo{journal}{Nucl. Phys.} \textbf{\bibinfo{volume}{A213}},
  \bibinfo{pages}{332} (\bibinfo{year}{1973}).

\bibitem[{\citenamefont{{Wilmes} et~al.}(1995)\citenamefont{{Wilmes}, {Mohr},
  {Atzrott}, {K{\" o}lle}, {Staudt}, {Mayer}, and {Hammer}}}]{wilmes95}
\bibinfo{author}{\bibfnamefont{S.}~\bibnamefont{{Wilmes}}},
  \bibinfo{author}{\bibfnamefont{P.}~\bibnamefont{{Mohr}}},
  \bibinfo{author}{\bibfnamefont{U.}~\bibnamefont{{Atzrott}}},
  \bibinfo{author}{\bibfnamefont{V.}~\bibnamefont{{K{\" o}lle}}},
  \bibinfo{author}{\bibfnamefont{G.}~\bibnamefont{{Staudt}}},
  \bibinfo{author}{\bibfnamefont{A.}~\bibnamefont{{Mayer}}}, \bibnamefont{and}
  \bibinfo{author}{\bibfnamefont{J.~W.} \bibnamefont{{Hammer}}},
  \bibinfo{journal}{\prc} \textbf{\bibinfo{volume}{52}}, \bibinfo{pages}{2823}
  (\bibinfo{year}{1995}).

\bibitem[{\citenamefont{{de Oliveira} et~al.}(1997)\citenamefont{{de Oliveira},
  {Coc}, {Aguer}, {Bogaert}, {Kiener}, {Lefebvre}, {Tatischeff}, {Thibaud},
  {Fortier}, {Maison} et~al.}}]{oliveira97}
\bibinfo{author}{\bibfnamefont{F.}~\bibnamefont{{de Oliveira}}},
  \bibinfo{author}{\bibfnamefont{A.}~\bibnamefont{{Coc}}},
  \bibinfo{author}{\bibfnamefont{P.}~\bibnamefont{{Aguer}}},
  \bibinfo{author}{\bibfnamefont{G.}~\bibnamefont{{Bogaert}}},
  \bibinfo{author}{\bibfnamefont{J.}~\bibnamefont{{Kiener}}},
  \bibinfo{author}{\bibfnamefont{A.}~\bibnamefont{{Lefebvre}}},
  \bibinfo{author}{\bibfnamefont{V.}~\bibnamefont{{Tatischeff}}},
  \bibinfo{author}{\bibfnamefont{J.-P.} \bibnamefont{{Thibaud}}},
  \bibinfo{author}{\bibfnamefont{S.}~\bibnamefont{{Fortier}}},
  \bibinfo{author}{\bibfnamefont{J.~M.} \bibnamefont{{Maison}}},
  \bibnamefont{et~al.}, \bibinfo{journal}{Phys. Rev. C}
  \textbf{\bibinfo{volume}{55}}, \bibinfo{pages}{3149(R)}
  (\bibinfo{year}{1997}).

\bibitem[{\citenamefont{Rolfs and Rodney}(1988)}]{rolfs88}
\bibinfo{author}{\bibfnamefont{C.~E.} \bibnamefont{Rolfs}} \bibnamefont{and}
  \bibinfo{author}{\bibfnamefont{W.~S.} \bibnamefont{Rodney}},
  \emph{\bibinfo{title}{Cauldrons in the Cosmos}} (\bibinfo{publisher}{The
  University of Chicago Press}, \bibinfo{address}{Chicago},
  \bibinfo{year}{1988}).

\bibitem[{\citenamefont{{Langanke} et~al.}(1986)\citenamefont{{Langanke},
  {Wiescher}, {Fowler}, and {G\"{o}rres}}}]{langanke86}
\bibinfo{author}{\bibfnamefont{K.}~\bibnamefont{{Langanke}}},
  \bibinfo{author}{\bibfnamefont{M.}~\bibnamefont{{Wiescher}}},
  \bibinfo{author}{\bibfnamefont{W.~A.} \bibnamefont{{Fowler}}},
  \bibnamefont{and}
  \bibinfo{author}{\bibfnamefont{J.}~\bibnamefont{{G\"{o}rres}}},
  \bibinfo{journal}{Astrophys. J.} \textbf{\bibinfo{volume}{301}},
  \bibinfo{pages}{629} (\bibinfo{year}{1986}).

\bibitem[{\citenamefont{{Vancraeynest}
  et~al.}(1998)\citenamefont{{Vancraeynest}, {Decrock}, {Gaelens}, {Huyse},
  {van Duppen}, {Bain}, {Davinson}, {Page}, {Shotter}, {Woods}
  et~al.}}]{vancraeynest98}
\bibinfo{author}{\bibfnamefont{G.}~\bibnamefont{{Vancraeynest}}},
  \bibinfo{author}{\bibfnamefont{P.}~\bibnamefont{{Decrock}}},
  \bibinfo{author}{\bibfnamefont{M.}~\bibnamefont{{Gaelens}}},
  \bibinfo{author}{\bibfnamefont{M.}~\bibnamefont{{Huyse}}},
  \bibinfo{author}{\bibfnamefont{P.}~\bibnamefont{{van Duppen}}},
  \bibinfo{author}{\bibfnamefont{C.~R.} \bibnamefont{{Bain}}},
  \bibinfo{author}{\bibfnamefont{T.}~\bibnamefont{{Davinson}}},
  \bibinfo{author}{\bibfnamefont{R.~D.} \bibnamefont{{Page}}},
  \bibinfo{author}{\bibfnamefont{A.~C.} \bibnamefont{{Shotter}}},
  \bibinfo{author}{\bibfnamefont{P.~J.} \bibnamefont{{Woods}}},
  \bibnamefont{et~al.}, \bibinfo{journal}{\prc} \textbf{\bibinfo{volume}{57}},
  \bibinfo{pages}{2711} (\bibinfo{year}{1998}).

\bibitem[{\citenamefont{{Coc} et~al.}(2000)\citenamefont{{Coc}, {Hernanz},
  {Jos{\' e}}, and {Thibaud}}}]{coc00}
\bibinfo{author}{\bibfnamefont{A.}~\bibnamefont{{Coc}}},
  \bibinfo{author}{\bibfnamefont{M.}~\bibnamefont{{Hernanz}}},
  \bibinfo{author}{\bibfnamefont{J.}~\bibnamefont{{Jos{\' e}}}},
  \bibnamefont{and} \bibinfo{author}{\bibfnamefont{J.-P.}
  \bibnamefont{{Thibaud}}}, \bibinfo{journal}{Astron. Astrophys.}
  \textbf{\bibinfo{volume}{357}}, \bibinfo{pages}{561} (\bibinfo{year}{2000}).

\bibitem[{\citenamefont{{Iliadis} et~al.}(2002)\citenamefont{{Iliadis},
  {Champagne}, {Jos{\' e}}, {Starrfield}, and {Tupper}}}]{iliadis02}
\bibinfo{author}{\bibfnamefont{C.}~\bibnamefont{{Iliadis}}},
  \bibinfo{author}{\bibfnamefont{A.}~\bibnamefont{{Champagne}}},
  \bibinfo{author}{\bibfnamefont{J.}~\bibnamefont{{Jos{\' e}}}},
  \bibinfo{author}{\bibfnamefont{S.}~\bibnamefont{{Starrfield}}},
  \bibnamefont{and} \bibinfo{author}{\bibfnamefont{P.}~\bibnamefont{{Tupper}}},
  \bibinfo{journal}{Astrophys. J. Suppl. Ser.} \textbf{\bibinfo{volume}{142}},
  \bibinfo{pages}{105} (\bibinfo{year}{2002}).

\end{thebibliography}
\end{document}